\documentclass{article}
\usepackage{tenor2025}
\usepackage{ifpdf}
\usepackage[english]{babel}
\usepackage{balance}
\usepackage{amsmath}
\usepackage{microtype}
\usepackage{booktabs}
\usepackage{multirow}
\usepackage{amsmath,amssymb}
\DeclareFontFamily{U}{musix}{}%
\DeclareFontShape{U}{musix}{m}{n}{%
  <-12>   musix11
  <12-15> musix13
  <15-18> musix16
  <18-23> musix20
  <23->   musix29
}{}%
\newcommand*\musix{\usefont{U}{musix}{m}{n}\selectfont}
\DeclareTextFontCommand{\textmusix}{\musix}

\newcommand*\doubleflat{\raisebox{.6ex}{\textmusix{3}}}
\newcommand*\doublesharp{\raisebox{.6ex}{\textmusix{5}}}

\def\papertitle{PAPER TEMPLATE FOR TENOR 2025}
\def\firstauthor{First author}
\def\secondauthor{Second author}
\def\thirdauthor{Third author}

\newif\ifpdf
\ifx\pdfoutput\relax
\else
   \ifcase\pdfoutput
      \pdffalse	
   \else
      \pdftrue
\fi

\ifpdf 
  \usepackage[pdftex,
    pdftitle={\papertitle},
    pdfauthor={\firstauthor, \secondauthor, \thirdauthor},
    bookmarksnumbered, 
    pdfstartview=XYZ 
   ]{hyperref}

  \usepackage[pdftex]{graphicx}
  \graphicspath{{./figures/}}
  \DeclareGraphicsExtensions{.pdf,.jpeg,.png}

  \usepackage[figure,table]{hypcap}

\else 
  \usepackage[dvips,
    bookmarksnumbered, 
    pdfstartview=XYZ 
  ]{hyperref}  

  \usepackage[dvips]{epsfig,graphicx}
  \graphicspath{{./figures/}}
  \DeclareGraphicsExtensions{.eps}

  \usepackage[figure,table]{hypcap}
\fi

\hypersetup{
    colorlinks,%
    citecolor=black,%
    filecolor=black,%
    linkcolor=black,%
    urlcolor=black
}

\title{EngravingGNN: A Hybrid Graph Neural Network \\ for End‐to‐End Piano Score Engraving}

 \threeauthors
   {\firstauthor} {Affiliation1 \\ %
     {\tt \href{mailto:author1@adomain.org}{author1@adomain.org}}}
   {\secondauthor} {Affiliation2 \\ %
     {\tt \href{mailto:author2@adomain.org}{author2@adomain.org}}}
   {\thirdauthor} { Affiliation3 \\ %
     {\tt \href{mailto:author3@adomain.org}{author3@adomain.org}}}

\oneauthor{Emmanouil Karystinaios$^{1}\thanks{* Equal contribution.}$ \hspace{0.3cm} Francesco Foscarin$^{2}$ \hspace{0.3cm} Gerhard Widmer$^{1,3}$}
{$^1$ Johannes Kepler University, Linz, Austria\\
$^2$ Moises Systems Inc., Salt Lake City, USA.\\ 
$^3$ LIT AI Lab, Linz Institute of Technology, Austria\\
{\tt\small firstname.lastname@jku.at}}

\begin{document}
\capstartfalse
\maketitle
\capstarttrue
\begin{abstract}
This paper focuses on automatic music engraving, i.e., the creation of a humanly-readable musical score from musical content. This step is fundamental for all applications that include a human player, but it remains a mostly unexplored topic in symbolic music processing. In this work, we formalize the problem as a collection of interdependent subtasks, and propose a unified graph neural network (GNN) framework that targets the case of piano music and quantized symbolic input. Our method employs a multi-task GNN to jointly predict voice connections, staff assignments, pitch spelling, key signature, stem direction, octave shifts, and clef signs. A dedicated postprocessing pipeline generates print-ready MusicXML/MEI outputs. Comprehensive evaluation on two diverse piano corpora (J-Pop and DCML Romantic) demonstrates that our unified model achieves good accuracy across all subtasks, compared to existing systems that only specialize in specific subtasks. These results indicate that a shared GNN encoder with lightweight task-specific decoders in a multi-task setting offers a scalable and effective solution for automatic music engraving.

\end{abstract}

\section{Introduction}\label{sec:intro}

The musical score encodes a variety of information that we can organize into two classes: the \textit{musical content}, which specifies the sounds that a musician should produce, and the \textit{score engraving}, i.e., the notation symbols which don't influence the music, but are employed for its graphical representation~\cite{foscarin2021data}. Examples of engraving information include staff separation, clefs, pitch spelling, grouping into chords and voices, octave shifts, and stem directions. A high-quality engraving is a fundamental component of a musical score as it allows musicians and musicologists to efficiently read it and get the composer’s intentions with precision. 

The situation changes significantly when we exclude the human component and consider music that must be played or composed by a machine, such as a sequencer. In this scenario, formats that only encode the musical content, like a MIDI-file, are much more efficient and easier to handle. This is visible in the field of music information retrieval (MIR) where the vast majority of systems, for example, automatic music transcription or automatic composition systems, work on MIDI-like representations.
This task of \textit{automatic music engraving} is then a necessary bridge to allow musicians to play computer-processed music.





\begin{figure}[t]
  \centering
  \includegraphics[width=\columnwidth]{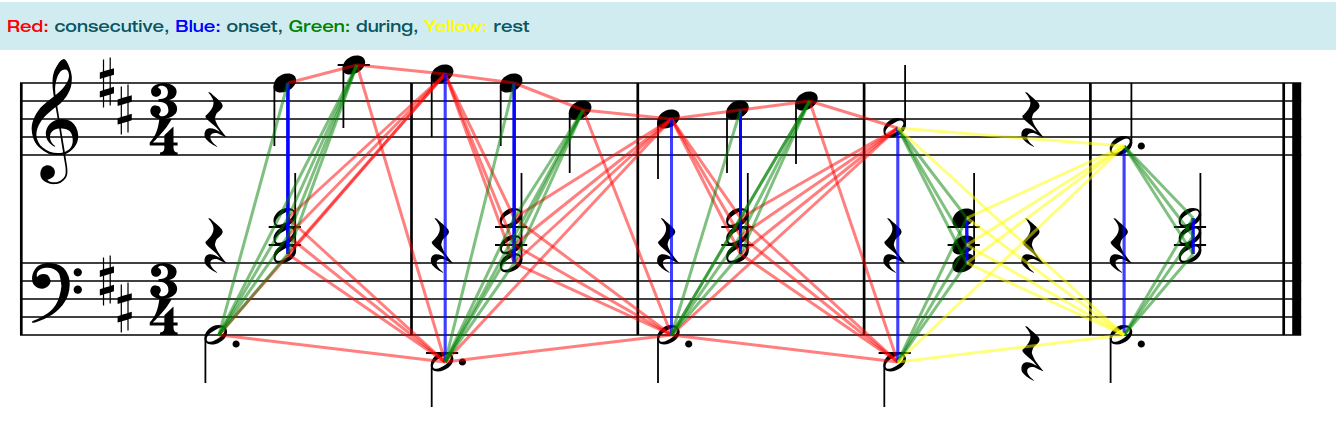}
  \caption{The score graph, nodes are notes and edges are marked in different color for each type.}
  \label{fig:graph}
\end{figure}



Some systems were recently proposed to address individual engraving subtasks such as pitch spelling and key estimation~\cite{pkspell, bouquillard2024engraving}, voice and staff separation~\cite{karystinaios2024voice}. We use the latter as a basis for our system, and improve it by extending it to more engraving tasks without reducing its performance.
Another related work is the one from Beyer and Dai~\cite{beyer2024end}, which targets in a single end-to-end framework the tasks of note quantization and music engraving. This is an interesting direction for automatic music transcription (AMT), but we keep our model more generic to be able to employ it in diverse scenarios (e.g., with a music generation system).

In this paper, we focus specifically on piano music and quantized inputs, introducing EngravingGNN: a unified framework for automatic score engraving. At its core, EngravingGNN employs a Hybrid-GNN encoder, a combination of heterogeneous graph convolutions and stacked GRU layers~\cite{karystinaios2024graphmuse}, to fuse relational note interactions with long-term temporal context. From this shared embedding, lightweight decoder heads simultaneously predict all of the following:

\begin{itemize}
  \item \emph{Voice edges} and \emph{staff labels}, organizing notes into independent streams on the appropriate stave,
  \item \emph{Pitch spelling} and \emph{key signature}, ensuring correct accidentals and tonal context,
  \item \emph{Stem direction} and \emph{octave shift} markings, adhering to engraving conventions for clarity,
  \item \emph{Symbolic duration} predicting dots, triplets, and notehead types (whole, half, eight, etc.),
  \item \emph{Clef signs}, producing the initial clefs and capturing mid‐piece clef changes when they occur.
\end{itemize}

In summary, our contributions are 3-fold:
\begin{enumerate}
  \item A end-to-end model network architecture for music engraving, with an \emph{Hybrid-GNN} encoder integrating heterogeneous graph convolutions with GRU stacks, plus 10 decoder heads for all core engraving tasks.
  \item A postprocessing and export pipeline yielding fully engraved MusicXML/MEI scores.
  \item Evaluation and comparison with existing systems on two piano corpora, with detailed metrics for voice, staff, pitch spelling, key, stem direction, symbolic duration, octave shifts, and clefs.
\end{enumerate}

\section{Methodology}\label{sec:method}
\begin{figure*}[t]
  \centering
  \includegraphics[width=\textwidth]{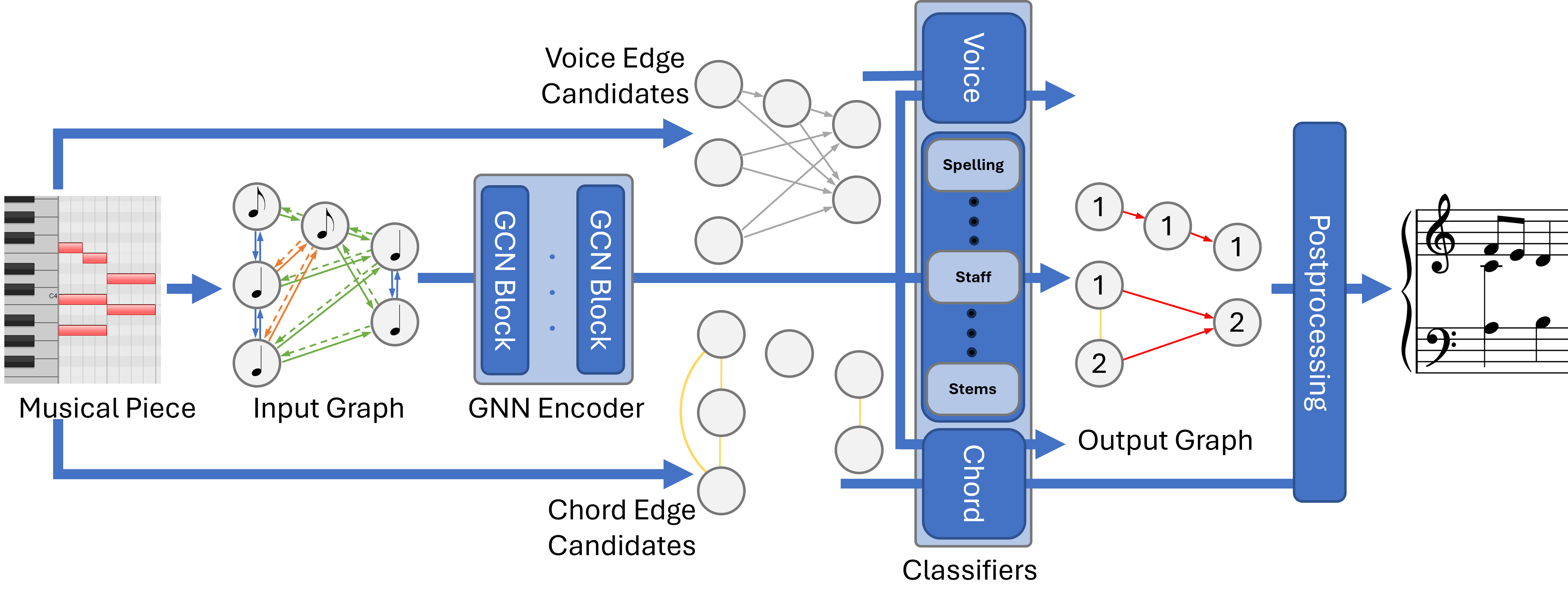}
  \caption{System overview: Input graph $\to$ GNN encoder $\to$ multi‐task decoders $\to$ postprocessing $\to$ engraved score.}
  \label{fig:model}
\end{figure*}

Figure~\ref{fig:model} provides a schematic overview of our approach. Starting from a quantized score, we construct an input graph, which is then encoded using a graph neural network (GNN). Multiple tasks are decoded in parallel from this encoded representation, followed by postprocessing and export of the final musical score.

\subsection{Input Graph Construction}\label{subsec:input_graph}
Given quantized notes, we create a directed heterogeneous graph $G_{in}=(V,E_{in},\mathcal{R}_{in})$:
\begin{itemize}
  \item \textbf{Nodes $V$}: one per note, with features:
    \begin{itemize}
      \item pitch class (12‐way one‐hot),
      \item octave (scalar in $[1,7]$),
      \item normalized duration $d\in(0,1)$ via $\tanh(\frac{\text{note }d}{\text{bar }d})$,
      \item onset fraction within bar and downbeat positions.
    \end{itemize}
  \item \textbf{Edge types $\mathcal{R}_{in}$}: onset, during, follow, silence, plus inverse edges, as in \cite{karystinaios2024voice}
\end{itemize}
In the current version of our system, grace notes are removed as they pose further challenges due to the lack of a quantized duration.
\subsection{Output Predictions}\label{subsec:output}

Our model produces a rich set of engraving annotations by formulating each attribute as either a node‐classification or an edge‐prediction task.  Given the GNN‐computed embedding \(\mathbf{h}_v\) for each note \(v\), and embeddings \((\mathbf{h}_u,\mathbf{h}_w)\) for each candidate note‐pair \((u,w)\), we jointly predict the following:

\begin{enumerate}
  \item \textbf{Voice edges.}  
    We treat voice assignment as an edge‐prediction problem over a restricted set of candidate pairs \(\Lambda_v\) as in \cite{karystinaios2024voice}.  Each candidate \((u,w)\in\Lambda_v\) satisfies that \(u\) and \(w\) lie in the same bar and \(\mathrm{offset}(u)\le\mathrm{onset}(w)\), ensuring temporal consistency.  A 2‐layer MLP, followed by a sigmoid function, takes the concatenated pair \([\mathbf{h}_u;\mathbf{h}_w]\) and outputs a probability \(p_{uw}\in(0,1)\) of \(w\) immediately following \(u\) in the same voice stream.
  
  \item \textbf{Staff label.}  
    Each note is assigned to one of two staves (upper or lower) via a binary node classifier.  A 2‐layer MLP, followed by a sigmoid function, maps \(\mathbf{h}_v\) to a probability \(s_v\).  At inference time, we threshold \(s_v\) at 0.5 to decide staff membership.  This separation is crucial for later beam grouping and rest insertion.
  
  \item \textbf{Pitch spelling.}  
    Correct notation requires exact accidentals.  We predict one of 35 classes per node (7 natural pitch classes \(\{A,B,C,D,E,F,G\}\), each with optional \(\sharp\), \(\flat\), \(\doublesharp \), \(\doubleflat \), or \(\natural\) sign).  A 2‐layer softmax MLP takes \(\mathbf{h}_v\) and outputs a distribution over these classes, trained via categorical cross‐entropy against MusicXML ground‐truth spellings.
  
  \item \textbf{Key signature.}  
    MusicXML represents key signatures as an integer in \([-7,+7]\) (number of fifth‐circle sharps or flats).  We add a sixteenth node‐classification head: a 2‐layer MLP with softmax over 15 classes.  Local key changes (a.k.a. modulations) occur in Romantic repertoire, so we predict this attribute per note and later smooth them within measures.
  
  \item \textbf{Stem direction.}  
    Stem orientation can be up or down or no-stem.  We model this as a ternary classification per note: Postprocessing enforces consistency within chords and aligns with beam direction.
  
  \item \textbf{Octave shift.}  
    Octave transposition indications (e.g.\ 8va, 8vb, 15ma) are common in piano music.  We predict a multiclass label per note among \{\texttt{none},\,8\texttt{va},\\ \,8\texttt{vb},\,15\texttt{ma}\} via a softmax MLP.  Consecutive notes sharing the same label are later merged into a single octave‐shift bracket in the score.
  
  \item \textbf{Clef sign.}  
    Piano scores occasionally employ clef changes (e.g.\ switch from G‐ to F‐clef on the upper staff).  We predict one of three classes: G‐, F‐, or C‐clef (this is rare in piano music, but we still support it for completeness) per note via a softmax head.  During export, we insert clef‐change tokens at the first note of each predicted clef region.
  
  \item \textbf{Symbolic duration.}  
    In addition to the quantized onset and offset, we must determine the proper notehead type (\(\tfrac{1}{16},\tfrac{1}{8},\tfrac{1}{4},\tfrac{1}{2},1,\dots\)), the number of augmentation dots, and any tuplet bracket (e.g.\ triplet) attributes.  We discretize “note‐type” into classes \(\{\texttt{whole},\tfrac12,\tfrac14,\tfrac18,\tfrac16,\dots\}\), “dots” into \(\{0,1,2,3\}\), and “tuplet” into common ratios \(\{1,3,5\}\) (i.e.\ no tuplet, triplet, quintuplet).  Each is predicted by its 2‐layer softmax MLP on \(\mathbf{h}_v\).  Postprocessing then groups adjacent notes with identical tuplet flags into a single bracket and assigns dot placement.
\end{enumerate}

All node‐classification heads input the same latent embedding from the GNN encoder.  Each head has its respective cross‐entropy loss, and the voice‐edge head uses binary cross‐entropy over \(\Lambda_v\). We sum these losses for end‐to‐end training.

\subsection{GNN Encoder}\label{subsec:encoder}

Our backbone encoder is a \emph{HybridGNN} that interleaves heterogeneous graph convolutions with recurrent GRU updates, combining the strengths of relational message‐passing and sequential modeling~\cite{karystinaios2024graphmuse}. By stacking these hybrid blocks, we capture both the rich, structured interactions between notes (via the graph convolutions) and the long‐term temporal persistencies across the piece (via the GRU stacks).

Formally, let each node \(u\) in our input graph \(G_{in}=(V,E_{in},\mathcal{R}_{in})\) have an initial feature vector \(\mathbf{h}_u^{(0)}\in\mathbb{R}^{d}\) encoding pitch class, octave, normalized duration, bar index, etc.  We apply \(L=3\) identical hybrid layers.  At layer \(l\), we first perform a heterogeneous GraphSAGE convolution \cite{hamilton2017representation}:

\[
\tilde{\mathbf{h}}_u^{(l)} 
= \sigma\Bigl(
  \mathbf{W}_0^{(l)}\,\mathbf{h}_u^{(l-1)} 
  \;+\;
  \sum_{r\in\mathcal{R}_{in}}
    \sum_{v \in \mathcal{N}_r(u)}
      \mathbf{W}_r^{(l)}\,\mathbf{h}_v^{(l-1)}
\Bigr),
\]

where \(\mathcal{N}_r(u)\) is the set of neighbors of \(u\) connected by relation type \(r\), and \(\{\mathbf{W}_r^{(l)}\}\) are relation‐specific weight matrices.  This step integrates multi‐relation context such as onset, during, follow, and silence edges.

In parallel, we feed the initial feature vectors \(\tilde{\mathbf{h}}_u^{(0)}\) into a gated recurrent unit (GRU) cell, the vectors are structured timely thanks to GraphMuse sampling process~\cite{karystinaios2024graphmuse}.

Because we stack one GRU per graph convolution layer, the network can accumulate and preserve information over long note sequences, learning temporal dependencies that pure GNN layers might lose.  The GRU update also allows each node to “remember” its evolving embedding across layers, reinforcing global structure.

After \(L=3\) such hybrid blocks, each note \(u\) has a final embedding \(\mathbf{h}_u^{(L)}\in\mathbb{R}^{256}\) that fuses local relational context with long‐range sequential patterns.  These embeddings serve as the shared inputs to all downstream decoders (voice‐edge prediction, staff label, pitch spelling, etc.), enabling coordinated multi‐task learning.

\subsection{Multi‐Task Decoders}\label{subsec:decoders}
Each task uses a lightweight 2‐layer MLP on $\mathbf{h}_u$ (or $[\mathbf{h}_u;\mathbf{h}_v]$ for edge tasks):
\begin{itemize}
  \item \textbf{Binary heads}: voice‐edge candidate and chord-edge candidates (sigmoid output + BCE loss).
  \item \textbf{Softmax heads}: staff, stem, pitch spelling, key signature, octave shift, clef, and symbolic duration (categorical cross‐entropy).
\end{itemize}

\subsection{Training}\label{subsec:training}
We train end‐to‐end, minimizing the (unweighted) sum of all task losses. We employ Adam optimizer (lr=1e‐3, weight decay $5\times10^{-4}$), 100 epochs. The train hyperparameters are taken from \cite{karystinaios2024voice}.

\subsection{Implementation Details} 

For the encoder, we use a hidden size of 256 for both the heterogeneous graph convolution outputs and the GRU hidden state.  Each convolutional layer is followed by a ReLU activation, and we apply 50\% dropout to the convolution output before it enters the GRU.  To stabilize training, layer normalization is applied both inside each GRU cell and between successive graph convolution blocks.  

\subsection{Postprocessing}\label{subsec:postprocessing}

A naive way to turn our voice‐edge probabilities into a voice separation would be to simply apply a threshold and keep all edges above it.  In practice, this can still produce three kinds of engraver’s nightmares:  
\begin{enumerate}
  \item Distinct voices collapse into one,  
  \item A single voice splits into multiple disjoint streams,  
  \item Notes that belong together in the same chord end up in different voices.  
\end{enumerate}
To guarantee musically valid output, we therefore follow prediction with a small postprocessing pipeline that enforces engraving constraints, as proposed in~\cite{karystinaios2024voice}.



\begin{figure}[t]
  \centering
  \includegraphics[width=\columnwidth]{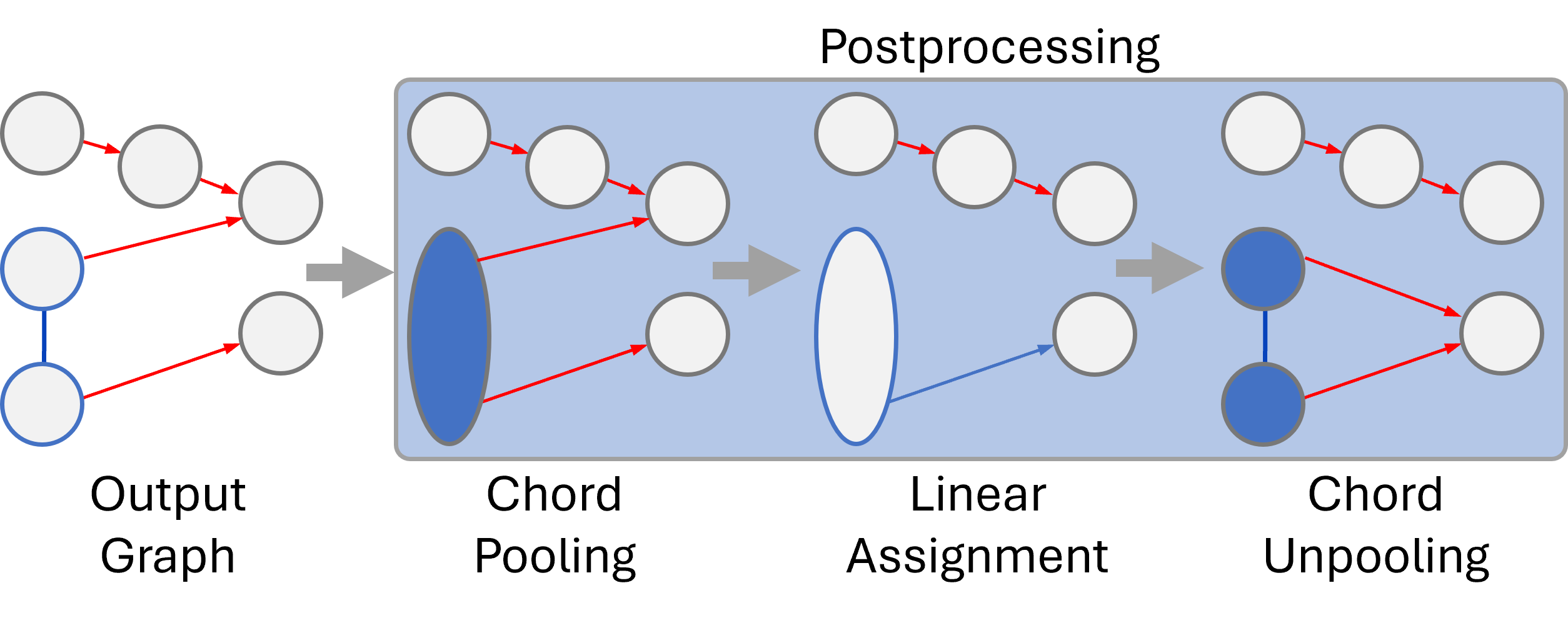}
  \caption{Postprocessing pipeline: chord pooling $\to$ voice assignment $\to$ unpooling $\to$ beaming \& rests.}
  \label{fig:postp}
\end{figure}

As shown in Figure~\ref{fig:postp}, we perform the following steps in order:
\begin{enumerate}
  \item \textbf{Chord pooling}: cluster synchronous notes (predicted chord edges) into virtual nodes.
  \item \textbf{Voice assignment}: Once chords are fused, the score reduces to purely monophonic streams, so we can solve a linear assignment via the Hungarian algorithm \cite{burkard1999linear} on pooled nodes.
  \item \textbf{Unpooling}: restore original notes, distributing assigned voices.  
  \item \textbf{Rest infilling}: insert rests to fill each bar per voice.
\end{enumerate}

We note that some node-level labels are altered by the chord-postprocessing to guarantee cohessiveness and proper engraving throughout predictions that need to carry the same label. Such labels include: symbolic duration, stem direction, and key signature. The logits of the notes that are pooled together for these tasks are averaged.
All other node-level attributes, i.e. octave shift, staff, pitch spelling, and clef sign, are independent of the post-processing operations.

\section{Experiments}\label{sec:experiments}
\subsection{Datasets}\label{subsec:datasets}

We evaluate on two piano corpora that span simple to highly complex repertoire. The \emph{J-Pop} set \cite{shibata2021non} (811 scores, 20\% held out) features straightforward pop arrangements—lower‐staff chords with simple upper‐staff melodies. In contrast, the \emph{DCML Romantic Corpus} \cite{dcml_piano_corpus} (393 scores, 20\% held out) includes 17th–20th-century works with virtuosic textures, cross-staff beaming, multiple overlapping voices, and complex engravings. We convert all sources to MusicXML (using MuseScore and Partitura \cite{partitura_mec}) and split each corpus into 80 \% train/validation and 20\% test sets, ensuring that our model’s ability to handle intricate engraving scenarios does not come at the cost of simpler pieces.

In line with Foscarin et al.~\cite{karystinaios2024voice}, we train EngravingGNN as a single, unified model on the combined training sets of both J‑Pop and DCML Romantic.


\subsection{Evaluation Metrics}\label{subsec:metrics}

We evaluate each engraving subtask with measures tailored to its output type. Voice separation is judged using the homophonic F1 metric from Hiramatsu et al.~\cite{hiramatsu2021joint}. Chord prediction is essentially a binary decision for each pair of synchronous notes: either they belong to the same chord or not.  We report the standard F1 score over all candidate chord‐edges.  

All remaining engraving attributes—staff assignment, stem direction, clef sign, key signature, pitch spelling, octave shift, and symbolic duration features (note head type, augmentation dots, tuplet bracket)—are treated as per‐note classification tasks. For each attribute, we compute the simple classification accuracy, i.e., the percentage of notes for which the predicted label exactly matches the ground truth. In the case of symbolic duration, which is decomposed into three separate predictions (note type, number of dots, tuplet flag), we measure both each individual head’s accuracy and the overall rate at which all three sub‐predictions are simultaneously correct.

All results are aggregated across the test set using micro‐averaging, ensuring that pieces with more notes or edges contribute proportionally to the final scores.

\subsection{Results}\label{subsec:results}

\begin{table*}[!htbp]
  \centering
  \small
  \begin{tabular}{lccccccccc}
    \toprule
    \multirow{2}{*}{Model} & \multirow{2}{*}{Voice F1} & \multirow{2}{*}{Chord F1} & Staff & Pitch & Key & Stem & Octave & Clef &  Sym. Dur.\\
      & & & Acc& Spelling Acc & Acc & Acc & Acc & Acc & Acc \\
    \midrule
    \multicolumn{8}{l}{\emph{J-Pop}}\\
    Spelling Model~\cite{karystinaios2024graphmuse} & - & - & - & \textbf{99.4} & \textbf{96.9} & - & - & - & -\\
    Shibata et al.\ \cite{shibata2021non} & 92.2 & & 92.8 & - & - & - & - & - & -\\
    Forscarin et al.\ \cite{karystinaios2024voice} & 96.6 & 94.9 & 96.3 & - & - & - & - & - & -\\
    \textbf{EngravingGNN}     & \textbf{96.8} & \textbf{96.9} & \textbf{97.6} & 96.3 & 80.6 & 85.1 & 99.9 & 96.2 & 87.8 \\
    \midrule
    \multicolumn{8}{l}{\emph{DCML Romantic}}\\    
    Spelling Model~\cite{karystinaios2024graphmuse} & - & - & - & 93.5 & 46.5 & - & - & - & -\\
    Shibata et al.\ \cite{shibata2021non} & 84.9 & & 88.5 & - & - & - & - & - & -\\
    Forscarin et al. \cite{karystinaios2024voice} & 89.9 & 79.5 & 91.0 & - & - & - & - & -  & -\\
    \textbf{EngravingGNN}     & \textbf{90.6} & \textbf{81.1} & \textbf{91.9} & 93.5 & \textbf{50.5} & 73.6 & 100 & 90.0 & 83.3 \\
    \bottomrule
  \end{tabular}
  \caption{Comparison of the test‐set results for our EngravingGNN model versus the Foscarin et al. ~\cite{karystinaios2024voice} and Shibata et al.\ \cite{shibata2021non} on the voice and staff tasks and versus the Spelling model of ~\cite{karystinaios2024graphmuse} for pitch spelling and Key signature prediction.}
  \label{tab:main_results}
\end{table*}

Table~\ref{tab:main_results} presents the test‐set performance of EngravingGNN alongside the baselines on both the J‑Pop and DCML Romantic corpora. On the J‑Pop dataset, EngravingGNN achieves a voice‐separation F1 of 96.8—slightly above the previous GNN‐only result (96.6) and well ahead of Shibata et al. (92.2). Chord grouping also improves to 96.9, and staff assignment climbs to 97.6\%.  While the specialized spelling model of Karystinaios and Widmer~\cite{karystinaios2024graphmuse} attains 99.4\% pitch and 96.9\% key‐signature accuracy, EngravingGNN’s integrated predictions still perform strongly at 96.3\% and 80.6\%, respectively. We believe that the reason for the degraded performance is the absence of data augmentation, which may be crucial for pitch spelling on a dataset without much key variability.
Stem direction (85.1\%), octave shifts (99.9\%), clef sign (96.2\%), and symbolic duration (87.8\%) likewise demonstrate high end‐to‐end engraving quality.

On the more demanding DCML Romantic corpus, which is characterized with attributes such as cross‑staff beaming, frequent modulations, and dense polyphony, EngravingGNN again leads against the other models. Voice F1 rises marginally to 90.6 (vs.\ 89.9), chord F1 to 81.1 (vs.\ 79.5), and staff accuracy to 91.9\% (vs.\ 91.0\%). Notably, it matches the standalone spelling model on pitch (93.5\%) and even surpasses it on key signature (50.5\% vs. 46.5\%), highlighting robust handling of chromatic passages and possible positive inter-task transfer. Stem direction (73.6\%) and symbolic duration (83.3\%) reflect the corpus’s notational complexity. Overall, EngravingGNN delivers consistent, high‐fidelity results across both straightforward and intricate piano repertoires. 

We remind the reader that the above results are from a single unified model, trained on the combined training sets of both J‑Pop and DCML Romantic. As Table~\ref{tab:main_results} shows, this shared model delivers strong, consistent performance on both test sets, demonstrating EngravingGNN’s ability to generalize across diverse piano repertoires without per‐dataset tuning.  

We do not evaluate against the model of Beyer and Dai~\cite{beyer2024end} since it would not be a fair comparison, as the quantization task they perform is very complex and diminishes the performance of the score engraving subpart of their system.

\subsection{Qualitative Analysis}\label{subsec:qualitative}
\begin{figure}[htbp]
  \centering
  \includegraphics[width=\columnwidth]{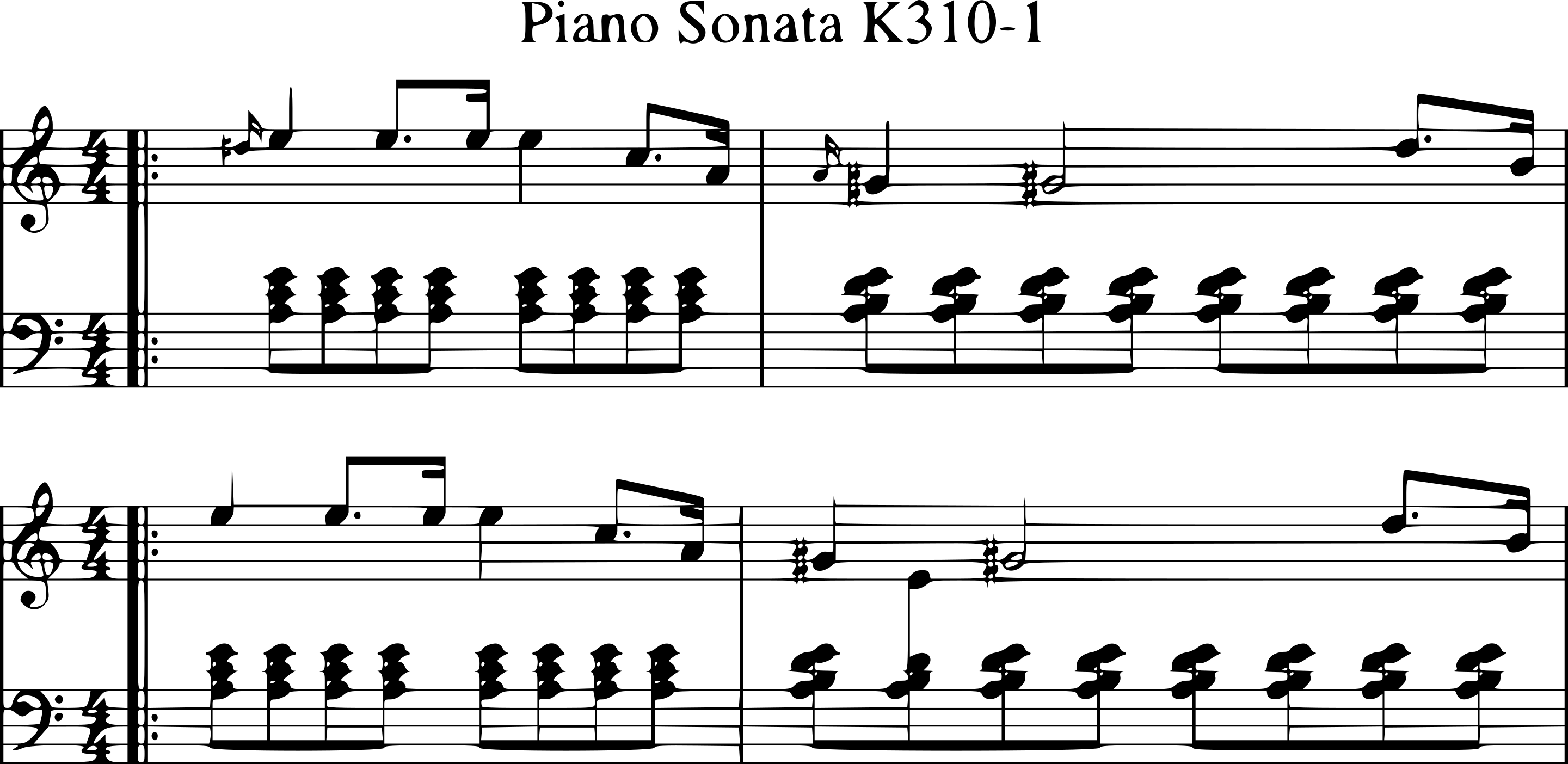}
  \caption{Engraving prediction on Sonata K310-1 by Mozart. Ground truth score on the top and EngravingGNN prediction on the bottom. Note that EngravingGNN, is not able to predict grace notes.
  }
  \label{fig:example}
\end{figure}

Figure~\ref{fig:example} illustrates our predictions on the first two bars of the first movements of Mozart's Sonata K310. Our method correctly captures symbolic duration, key, voices, staff, stem-direction, and spelling for each single note. We can also see a wrong prediction on the second half of the first beat, where the staff of an E which causes a cross staff assignment.

\section{Discussion and Limitations}\label{sec:discussion}

\subsection{Treating Tied Notes}

In its current version, our framework does not handle ties across multiple noteheads. In traditional engravings, ties connect notes across bars, across strong beats for better readability, or to form durations that cannot be expressed by one symbol alone. Since EngravingGNN’s duration head predicts only one symbol per input duration, it cannot generate the necessary tie‐edges or composite duration encodings. For this, autoregressive systems like the one of Bayer and Dai~\cite{beyer2024end}, which are naturally able to handle the ``one-to-many'' case, have a clear advantage. We believe that adding an autoregressive routine for the notehead task would be the way to improve in this direction.

\subsection{Including Grace Notes}

Our system does not currently support grace notes because grace notes lack quantized durations. However, grace notes are of crucial importance in many Baroque, Classical, and jazz passages, as they convey stylistic nuance. Proper engraving of grace notes would require flagging them in the input graph, linking them to their principal notes, and rendering them with small‐noteheads with appropriate slurs. Extending EngravingGNN to handle these ornaments is not trivial but it would be a key step toward fully authentic and stylistically complete piano scores.  

\subsection{Postprocessing}

While our postprocessing pipeline successfully pools chords, assigns voices, groups beams, and infills rests to produce valid MusicXML/MEI output, several limitations remain. First, symbolic‐duration handling is limited to single‐symbol durations and fixed tuplet brackets, so complex ties, composite rhythms, or wrong predictions can cause rendering errors in symbolic music visualizers. Second, global attributes such as key signature may exhibit measure‐to‐measure discontinuities, as we apply predictions per note without sequence‐level smoothing. Finally, our beam grouping and rest‐infilling steps rely on deterministic, rule‐based algorithms rather than learned patterns, which can result in suboptimal engraving choices in corner‐case scenarios (e.g.\ irregular beaming across voices or nonstandard rhythmic groupings).  

\subsection{Training and Optimization}

Hyperparameters were selected from prior work~\cite{karystinaios2024voice}. We did not perform an extensive hyperparameter search over learning rates, layer depths, or batch sizes was conducted. Consequently, while our off‐the‐shelf settings are enough for our goal of showcasing the potential of our approach, further gains may be possible through systematic hyperparameter search.

\section{Conclusion and Future Work}\label{sec:conclusion}

In this paper, we have presented EngravingGNN, a unified Hybrid‐GNN framework that, with a single encoder and multi‐head decoders, simultaneously predicts all essential piano engraving attributes, such as voice edges, staff assignment, pitch spelling, key signature, stem direction, octave shifts, and clef changes. After post-processing, our system generates ready‐to‐print MusicXML/MEI scores directly from quantized input, significantly reducing the manual effort traditionally required for high‐quality engraving.

Despite these advances, our current model assumes each note’s duration fits a single symbol in our predefined vocabulary, and does not handle tied notes or composite rhythms that span multiple noteheads or measures. It also leaves layout optimization, system breaks, staff spacing, and beam slopes, to external tools. Future work will extend the decoder to detect and represent ties and composite durations. Furthermore, we aim to explore end‐to‐end training with differentiable rendering feedback (e.g., via MEI/Verovio) to further close the gap between automated and professional human engraving.  

\section{Acknowledgements}
This work was supported by the European Research Council (ERC) under Horizon 2020 grant \#101019375 “Whither Music?”.

\balance
\bibliography{tenor2025-bib-template}

\end{document}